\documentclass[11pt]{JHEP3}
\usepackage{graphicx}
\usepackage{amsfonts}
\usepackage{epsfig}
\usepackage{amssymb}
\usepackage{amsmath}
\usepackage{epstopdf}
\newcommand{\be}{\begin{eqnarray}}
\newcommand{\ee}{\end{eqnarray}}
\newcommand{\bi}{\begin{itemize}}
\newcommand{\ei}{\end{itemize}}

\newcommand{\ba}{\begin{eqnarray}}
\newcommand{\ea}{\end{eqnarray}}

\title{Counting defects with the two-point correlator}

\author{Arttu Rajantie\\Theoretical Physics, Blackett Laboratory,
 Imperial College,  London SW7 2AZ, United Kingdom\\E-mail:  \email{a.rajantie@imperial.ac.uk}}
\author{Anders Tranberg\\Helsinki Institute of Physics, P.O.Box 41, FIN-00014 Helsinki, Finland and \\Department of Physical Sciences, University of Oulu, FIN-90014, Oulu, Finland\\E-mail: \email{anders.tranberg@helsinki.fi}}

\abstract{
We study how topological defects manifest themselves in the equal-time two-point field correlator. We consider a scalar field with $Z_2$ symmetry in 1, 2 and 3 spatial dimensions, allowing for kinks, domain lines and domain walls, respectively. Using numerical lattice simulations, we find that in any number of dimensions, the correlator in momentum space is 
to a very good approximation the product of two factors, one describing the spatial distribution of the defects and the other describing the defect shape. When the defects are produced by the
Kibble mechanism, the former has a universal form as a function of $k/n$, which we determine
numerically. This signature makes it possible to determine the kink density from the field correlator without having to resort to the Gaussian approximation. This is essential when studying field dynamics with methods relying only on correlators (Schwinger-Dyson, 2PI).
}

\keywords{Defects, out-of-equilibrium, field theory, real-time}

\preprint{Imperial/TP/10/AR/01, HIP-2010-6/TH}

\begin{document}


\section{Introduction\label{sec:introduction}}

Spontaneously broken global and local symmetries play an important role in many physical systems,
from high energy physics to condensed matter. In many cases, the symmetry breaking pattern is 
topologically non-trivial, giving rise to a possibility of topological defects, non-linear 
objects which are stable because of their topology. Examples of these include domain walls in 
ferromagnets, vortices in superfluids and superconductors, and magnetic monopoles and cosmic 
strings in high energy physics and cosmology.

In general, spontaneously broken symmetries are restored at sufficiently high temperatures. As 
the system cools down, it undergoes a phase transition into the broken phase. When this happens,
the system locally picks out one of the possible vacuum states. This choice is dynamical and 
random, and would generally have different outcome in different patches separated by more than 
one correlation length. This phenomenon, known as the Kibble mechanism, results in the creation 
of topological defects~\cite{kibble,Zurek:1985qw,Rajantie:2001ps}. It has been studied in detail in many experimental setups~%
\cite{Chuang:1991zz,Bowick:1992rz,Baeuerle:1996zz,Ruutu:1995qz,Digal:1998ak,%
Ducci,Casado,Maniv,Kirtley,Monaco}, 
and it may have also taken place in the early universe, producing domain walls, cosmic strings or 
magnetic monopoles.

The number of defects created in such a transition depends on the dynamics of the model, the 
nature of the symmetry and the cooling rate, which in the cosmological context is determined by 
the Hubble rate $H$.
This dependence can be estimated by considering the critical behaviour the theory and expressed
in terms of its critical exponents~\cite{Zurek:1985qw}, and these estimates have been shown to 
be accurate in some condensed matter experiments~\cite{Ducci,Casado} and lattice field theory 
simulations~\cite{Antunes:1996na,Laguna:1996pv,Laguna:1997cf,Antunes:1998rz}. In these 
simulations, the classical equations of motion are 
solved numerically on a spatial lattice, and the number of defects was determined by identifying them 
in the final field configuration and counting them. 
This approach is fully non-perturbative and can be used also in gauge field theories~\cite{Yates:1998kx,Hindmarsh:2000kd}, but
it cannot incorporate quantum mechanical effects and even in a classical theory, any noise such as thermal fluctuations can make the defects hard to distinguish, and the counting ill defined.

Because of these limitations, it would be advantageous to study defect formation from
first principles using other techniques that would give a more detailed understanding of the 
process and would be valid also in quantum theory. There have been many attempts to do this
using various techniques ranging from a linear 
approximation~\cite{Gill:1994ye,Karra:1996xf,Lythe,Uhlmann} 
to Schwinger-Dyson 
equations, Hartree approximation~\cite{Boyanovsky:1993pf,Antunes:1996na} 
and the 2PI formalism~\cite{Stephens:1998sm}. 

A common feature in these alternative approaches is that one does not
have access to individual field configurations, and therefore direct counting of defects is not 
possible. Instead, the dynamical variables are correlation functions. This raises the question of 
how to determine the number of defects from the field correlator. If one assumes that the field 
ensemble is Gaussian and identifies zeros of the field with defects, one can derive a simple 
expression for their density in terms of derivatives of the two-point 
correlator~\cite{Halperin,Liu,Ibaceta,Stephens:1998sm}. 
However, because defects are non-linear 
objects, their presence makes the field ensemble non-Gaussian, and therefore this assumption is 
not justified. Indeed, numerical simulations have shown that this approach does not 
work~\cite{defects2PI}.

In this paper, we show how kinks, domain lines and walls in 1, 2 and 3 spatial dimensions 
respectively, manifest themselves in the equal time two-point correlator in 
momentum space at a non-linear level, without relying on the Gaussian assumption. 
This is an extension of previous work~\cite{defects2PI}, in which we showed
that the 2PI formalism is unable to 
reproduce the classical signatures of global defects, at least in a $1/N$ expansion to 
next-to-leading order. 

We will first set up a simple model describing kinks, domain walls and domain lines (section \ref{sec:setup}), and then 
derive the signatures in the two-point function which we will be looking for (section 
\ref{sec:correlator}). We then perform sample lattice simulations in both 1, 2 and 3 spatial 
dimensions (section \ref{sec:numerics}), and demonstrate the signatures in practice, calibrating 
the net defect density the a benchmark density (section \ref{sec:results}). We conclude in 
section \ref{sec:conclusion}.


\section{\label{sec:setup}Setup, model and defects}

We consider the action of a real scalar field $\phi$ in $D$ spatial dimensions
\be
S=\int d^Dx\,dt\,\left(\frac{1}{2}\partial_{\mu}\phi\partial^{\mu}\phi-V(\phi)
\right),
\ee
with a potential $V(\phi)$, so that the theory has $Z_2$ symmetry ($\phi\leftrightarrow -\phi$),
\be
\label{equ:Vbroken}
V(\phi)=
-\frac{1}{2}\mu^{2}\phi^2+\frac{\lambda}{24}\phi^{4},
\ee
 The parameters and fields have energy dimensions $[\phi]=(D-1)/2$, $[\mu]=1$, $[\lambda]=3-D$.  The equation of motion reads,
\be
\label{equ:eom}
\partial_{t}^{2}\phi(x,t)+\Gamma\,\partial_{t}\phi(x,t)-\partial_{x}^{2}\phi(x,t)-\mu^{2}\phi(x,t)+\frac{\lambda}{6}\phi^3(x,t)=0.
\ee
and we have added a small damping term $\Gamma\partial_{t}\phi$ ($[\Gamma]=1$). This will drive the system from a high temperature initial state to a cold, near-vacuum final state (see also section \ref{sec:numerics}).

The model has two degenerate vacua at $\phi_{0}=\pm
v=\pm\sqrt{6/\lambda}\,\mu$, and there are
topological defects, kinks ($D=1$), domain lines ($D=2$), domain walls ($D=3$), which interpolate
between them. The classical kink solution is 
\be
\label{equ:kinkprofile}
\phi_{\rm kink}(x)=v\tanh\frac{x}{d},
\ee
where $d=\sqrt{2}/\mu$ is the kink thickness. Domain lines and domain walls are extensions in 1 and 2 dimensions of kinks, possibly with curvature on a length scale which we will assume to be much larger than $d$. 

In the following, we consider a situation in which a network of defects has been formed in a symmetry breaking phase transition by the Kibble mechanism.
In this case, the defect network has one characteristic length scale, which is determined by the field correlation length at the time of
the transition. It corresponds to the typical distance between defects or, equivalently, the number density $n$ of defects.


\section{The correlator ansatz in 1, 2 and 3 dimensions\label{sec:correlator}}

Our basic observable is the two-point correlator 
\be
G(r=|x-y|)=\langle\phi(x)\phi(y)\rangle,
\ee
assumed here to be homogeneous and isotropic, and therefore a function of the relative position $r$ only. The only two scales in the system are the kink thickness $d$ (or $\mu$), and the defect density $n$\footnote{Related in a complicated way to the damping rate $\Gamma$.}. In general, we would expect the position and momentum space correlators to be some general functions $G(r,d,n)$ and $G(k,d,n)$. In \cite{defects2PI}, we argued that in 1D, in the limit of $d=0$, and if kinks are distributed at random with average density $n$, the correlator should have the form
\be
G(r,n)=v^2e^{-2nr},\qquad G(k,n)=\frac{v^2}{2n}\frac{2}{1+(k/2n)^2}.
\ee
If the kinks all have the same smooth profile $\phi_{\rm kink}(x)$, this modifies the correlator by a multiplicative factor,
\be
\label{kinksig}
G(k,d,n)=\frac{v^2}{2n}\frac{2}{1+(k/2n)^2}\frac{k^2}{4}|\phi_{\rm kink}(k)|^2,
\ee
where $\phi_{\rm kink}(k)$ is the Fourier transform of the kink profile (\ref{equ:kinkprofile}), 
\be
\label{kinksolution}
\phi_{\rm kink}(k)=
{\frac{2iv}{k}}\frac{\frac{1}{2}\pi kd}{\sinh \frac{1}{2}\pi kd}.
\ee
Therefore we should find
\be
\label{equ:kinksign}
G(k,d,n)=\frac{v^2}{2n}\frac{2}{1+(k/2n)^2}\left(
\frac{\frac{1}{2}\pi kd}{\sinh\left(\frac{1}{2}\pi kd\right)}\right)^2
.
\ee
Since $d$ and $v$ are known\footnote{Strictly speaking, these are the vacuum values, but we will see that they are sufficient. We will comment on finite temperature effects below.}, this expression
allows for a one-parameter fit in $n$. More generally, the field correlator factorises as
\be
G(k,d,n)=\frac{v^2}{n}G_{\rm corr}(k/n)G_{\rm kink}(kd),
\ee
where $G_{\rm corr}(k/n)$ describes the spatial distribution of the kinks and $G_{\rm kink}(kd)$ 
the kink shape.
In particular, when the kinks are formed by the Kibble mechanism, the only relevant scale is 
their density, and therefore $G_{\rm corr}$ is a function of the dimensionless combination $k/n$ only.

We will generalise this procedure and instead make the ansatz 
that the field correlator factorises in the same way in higher dimensions,
\be
G(k,d,n)=\frac{v^2}{n^D}G_{\rm corr}^D(k/n)G_{\rm kink}(kd).
\ee
Several comments are in order at this point. Firstly, the $d$ and $n$ dependence are separated, and apart from the trivial scaling $n^{-D}$ coming from the measure in the Fourier transform, the correlator only depends on the dimensionless quantities $kd$ and $k/n$. Also, we can obtain the
kink correlation function, as
\be
G_{\rm corr}(k/n)\rightarrow\frac{n^D}{v^2}\frac{G(k,d,n)}{G_{\rm kink}\left(kd\right)}.
\label{eq:rescaled}
\ee
The Kibble mechanism predicts that this is a universal function of the ratio $k/n$ only,
and therefore if we measure this for different parameters and different cooling rates, the
results should coincide. We will use numerical simulations to
demonstrate that this is indeed the case, and that 
to a very good approximation, the universal function (in coordinate space) is of the form (for 
some $a_i,b_i$)
\be
G_{\rm corr}(nr)=\left(a_1e^{-a_2 (nr)^2}+b_1e^{-b_2nr}\right), 
\label{eq:universalform}
\ee
which in D-dimensional momentum space becomes (for some other $\alpha_i,\beta_i$)
\be
G_{\rm corr}^D\left(k/n\right)=\left(\alpha_1\, e^{-\alpha_2\,(k/n)^2}+\frac{\beta_1}{\left[1+\beta_2(k/n)^2\right]^{(D+1)/2}}\right).
\label{eq:gaussexp}
\ee
We use our numerical results to determine the parameters $\alpha_i$ and $\beta_i$. After this, the form of the correlator is fixed, and
the only remaining free parameter is $n$ which sets the scale. The defect density $n$ can 
therefore
be determined directly from the 
correlator using a simple one-parameter fit.


\section{Numerical procedure\label{sec:numerics}}

To carry out the numerical calculation, we discretised the equation of motion (\ref{equ:eom}) 
on a lattice of spacing $a$ and using a leapfrog algorithm with time step $a\delta t$ for the 
time derivative. Input parameters are lattice size $n_x^D$, $\delta t$, and the dimensionless 
$a\mu$, $a\Gamma$, $a^{D-3}\lambda$. 

We choose the initial conditions at time $t=0$ to mimic the quantum vacuum state corresponding to 
the potential
\be
\label{equ:Vini}
V_{\rm ini}(\phi)=\frac{1}{2}\mu^2\phi^2.
\ee
Because this is a free theory, the equal-time quantum two-point functions of the field $\phi$ and 
its canonical momentum 
$\pi=\partial_{t}\phi$ are simply
\ba
\label{equ:initcond}
\langle
\phi(k)\phi(q)\rangle=(2\pi)\delta(k+q)
\frac{1}{2\omega_k},\quad
\langle
\pi(k)\pi(q)\rangle=(2\pi)\delta(k+q)
\frac{\omega_k}{2},\quad
\langle
\phi(k)\pi(q)\rangle=0,
\ea
where  $\omega_k=\sqrt{k^2+\mu^2}$. Our initial condition is realised by a Gaussian ensemble of field configurations which has these same two-point functions~\cite{Khlebnikov:1996mc}. 
We initialise all modes, and although this corresponds to a divergent energy in the continuum limit, at finite lattice spacing it is just a specific choice. In our case, it is not crucial how well these initial conditions reproduce the actual initial quantum state, since we are only interested in the classical dynamics. For our purpose, it is sufficient that it corresponds to a state with unbroken symmetry\footnote{Other possible choices could be a Bose-Einstein distribution (finite UV energy), or a classical thermal distribution (divergent UV energy). We found that it made very little difference to the final number of defects, as long as the initial energy density was high enough. See for instance \cite{Smit:2001qa,GarciaBellido:2002aj,Arrizabalaga:2004iw} for studies of different initial condition prescriptions in the context of symmetry breaking and thermalisation. }.

The classical equation of motion (\ref{equ:eom}) 
allows us to rescale the coupling $\lambda$ to unity, suggesting that only the dimensionless ratio $\lambda/\mu^2$ plays a role. However,
the initial conditions (\ref{equ:initcond}) remove this freedom. 

We solve the time evolution of the system using the classical equation of motion (\ref{equ:eom}) for
a large number of initial configurations that are drawn from the distribution specified by Eq.~(\ref{equ:initcond}). Because of the inclusion of a damping term $\Gamma$, the system will lose energy and cool down to a state close to the vacuum. However, as we are going through a symmetry breaking transition, topological defects will be created. At the end of the evolution, these will be ``frozen'' in, possibly with very small residual thermal fluctuation superposed. The correlator at this final time is our observable, from which we will extract the number of defects.


\section{Calibration and results\label{sec:results}}

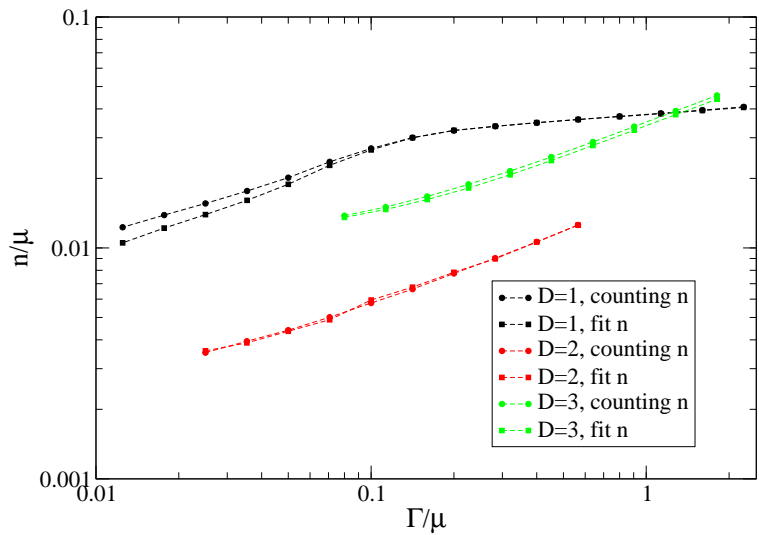
\EPSFIGURE{./Pictures/n_allD.eps,width=10cm,clip}{The ``counting'' defect density $n$ for different $\Gamma$ in $D=1,2,3$ dimensions (circles). Superposed, the density $n$ determined from a fit of the field correlator to our ansatz (\ref{eq:gaussexp}).
\label{fig:n_allD} }

To determine the parameters $\alpha_i$, $\beta_i$ in (\ref{eq:gaussexp}) from the numerical
data, we need to have an independent measurement of the defect density $n$.
We do this by  direct counting of zeros. We count, configuration by configuration,  all lattice points $N_{\rm points}$ where the field value is in some range $[-\theta,\theta]$ around $\phi=0$. We divide by the width of the kink profile in that range,
\be
\tanh[\Delta x/d]^2<\theta,\qquad d=\sqrt{2}/(a\mu).
\ee
giving for instance
\be
\theta=0.1,\rightarrow \Delta x(\theta)=\pm 0.327 d,\qquad \theta=0.05,\rightarrow \Delta x(\theta)=\pm 0.227 d,\\
\theta=0.025, \rightarrow \Delta x(\theta)=\pm 0.159d,\qquad \theta=0.0125\rightarrow \Delta x(\theta)=\pm 0.112 d.
\ee
The number of kinks/length of wall is then the number of lattice points divided by this length, and the density (in lattice units) is 
\be
\label{eq:countingn}
n(\theta)=\frac{N_{\rm points}}{2\Delta x(\theta) n_x^D N_{\rm configs}}.
\ee
This we consider to be the benchmark defect density, at least for late times. We checked that except at very early time, when the walls are not well separated, the kink/wall density is practically independent of the threshold $\theta$. This conclusion holds for $D=1, 2, 3$. The counting density as a function of damping rate is shown in Fig.~\ref{fig:n_allD}.

For a given $\Gamma$, we calculate this "counting" density, and use it to rescale the correlator as in Eq.~(\ref{eq:rescaled}). If the resulting curves coincide, we will have shown that 1) the original correlator is separable in $kd$ and $k/n$, 2) the dependence on $n$ is through $nr$ only 3) $n$ is in fact the "counting" $n$ (\ref{eq:countingn}), up to an overall calibration and 4) the rescaled curve is $G_{\rm corr}$, which one can then attempt to approximate. Note that since the density ranges over orders of magnitude (Fig.~\ref{fig:n_allD}), the curves certainly do not coincide without rescaling.

\DOUBLEFIGURE
{./Pictures/rescaled1D.eps,width=7cm,clip}
{./Pictures/gaussexp1D.eps,width=7cm,clip}
{
\label{fig:1D1}
The rescaled propagators (\protect\ref{eq:rescaled}) in $D=1$ for different $\Gamma$.
}
{
\label{fig:1D2}
One rescaled propagator ($\Gamma/\mu=0.1$) with high statistics. Superposed, the fitted form (\protect\ref{eq:universalform}), as well as the Gaussian and exponential components separately.
}

Fig.~\ref{fig:1D1} shows the rescaled curves for various $\Gamma$ for simulations in $D=1$ spatial dimensions. We use a lattice of $2^{17}=131072$ points, $a\mu=0.1$, and $a^2\lambda=0.6$.  We have $v^2=0.1$, $d=\sqrt{2}/a\mu\simeq 14.14$. We consider a range of damping rates $\frac{\Gamma}{\mu}=0.05-3.2$. The curves agree very well indeed. In Fig.~\ref{fig:1D2} we show a single rescaled curve with improved statistics (16 times as many configurations) and the overlaid fit of the anticipated form (\ref{eq:universalform}). The fit is quite convincing, although at intermediate scales, there is an oscillating feature which we choose not to account for in our description. The fit parameters $\alpha_i$, $\beta_i$ are given in Table~\ref{tab:fitparams}.

\TABLE{
\begin{tabular}{ccccc}
$D$ & $\alpha_1$ & $\alpha_2$ & $\beta_1$ & $\beta_2$\cr
\hline
$1$ & $0.384\pm 0.001$ & $0.0356\pm 0.0001$ & $0.180\pm 0.001$ & $0.0447\pm 0.0004$\cr
$2$ & $0.725\pm 0.008$ & $0.1039\pm 0.0006$ & $0.274\pm 0.010$ & $0.097\pm 0.003$\cr
$3$ & $1.73\pm 0.01$ & $0.1654\pm 0.0004$ & $0.406\pm 0.013$ & $0.110\pm 0.003$
\end{tabular}
\label{tab:fitparams}
\caption{The fit parameters in (\protect\ref{eq:gaussexp}) determined from the numerical data.}
}

\DOUBLEFIGURE
{./Pictures/rescaled2D.eps,width=7cm,clip}
{./Pictures/gaussexp2D.eps,width=7cm,clip}{The rescaled propagators (\protect\ref{eq:rescaled}) in $D=2$ for different $\Gamma$.
\label{fig:2D1}}
{One rescaled propagator ($\Gamma/\mu=0.1$) with high statistics. Superposed, the fitted form (\protect\ref{eq:universalform}), as well as the Gaussian and exponential components separately.
\label{fig:2D2}}

Similarly, Fig.~\ref{fig:2D1} shows the same method applied to simulations in $D=2$.  We use a $2048^2$ lattice with $a\mu=0.4$, $a\lambda=0.6$. We have $av^2=1.6$, $d=\sqrt{2}/a\mu\simeq 3.535$, and we use $\frac{\Gamma}{\mu}=0.025-0.4$. Again, the rescaled curves agree very well, and again (Fig.~\ref{fig:2D2}) the simple form (\ref{eq:universalform}) is an excellent fit. 
The fit parameters $\alpha_i$, $\beta_i$ are given in Table~\ref{tab:fitparams}.

\DOUBLEFIGURE
{./Pictures/rescaled3D.eps,width=7cm,clip}
{./Pictures/gaussexp3D.eps,width=7cm,clip}
{The rescaled propagators (\protect\ref{eq:rescaled}) in $D=3$ for different $\Gamma$.
\label{fig:3D1}}
{One rescaled propagator ($\Gamma/\mu=0.1$) with high statistics. Superposed, the fitted form (\protect\ref{eq:universalform}), as well as the Gaussian and exponential components separately. 
\label{fig:3D2}}

Finally, Figs.~\ref{fig:3D1} and \ref{fig:3D2} show results in $D=3$, with a fit to the appropriate version of (\ref{eq:universalform}). We use a $256^3$ lattice, $a\mu=1.0$, $\lambda=0.6$ to give $a^2v^2=10$, $d=\sqrt{2}$. $\frac{\Gamma}{\mu}=0.08-0.64$. The match is again very good, and we fidn
the fit parameters $\alpha_i$, $\beta_i$ given in Table~\ref{tab:fitparams}.

For the very large $\Gamma$ curve (magenta), the agreement in the UV is not perfect, a result of having a large defect density, $n\simeq 0.08$, $nd\simeq 0.11$. This means that 11 percent of all lattice points are in the core of a wall, and it is therefore not surprising that our separation of scales does not hold. The wall profile is no longer a simple $\tanh$ of $r/d$. But in the IR the agreement is still good, and for all smaller damping rates even the UV performs well. 

With the parameters $\alpha_i$, $\beta_i$ in (\ref{eq:gaussexp}) known, it is possible to
determine the defect density $n$ from the correlator by a simple fit to (\ref{eq:gaussexp}), 
keeping only $n$ as the free parameter. 
We demonstrate this by calculating $n$ as a function of $\Gamma$ in this way.
Fig.~\ref{fig:n_allD} shows the ``counting'' densities $n$ and the ``fit'' $n$, resulting from this procedure. The agreement is excellent (within 2 percent), except for very small damping rates $\Gamma$ in 1D, where presumably residual thermal noise introduces the discrepancy. Note that for each $D$, the calibration was performed with a separate, high-statistics, simulation at one particular value of $\Gamma$. These are not included in Fig.~\ref{fig:n_allD}, and so although because of the successful rescalings Figs.~\ref{fig:1D1}, \ref{fig:2D1}, \ref{fig:3D1} the match could be anticipated, it is a non-trivial result.


\section{Conclusion\label{sec:conclusion}}

We have shown that the equal-time momentum space two-point field correlator
in a defect configuration produced by the Kibble mechanism 
separates in an UV part encoding the kink profile (in terms of $d$) and an IR part encoding the distribution of kinks (in terms of the density $n$). 
We carried out numerical simulations to obtain
a very good approximation to the exact form, which turned out to be the sum of a Gaussian and a position-space exponential. This is a somewhat surprising result, since for instance assuming randomly distributed kinks in $D=1$ yields only the exponential. Furthermore, we found that the IR part of the correlator only depends on the defect density and scales
with it in a simple way. This is a highly non-trivial result and allows us to determine the defect
density with a one-parameter fit to the field correlator.

This is important because direct counting of defects, which usually works well in classical field theory simulations, is not a meaningful procedure in full quantum field theory because the
state of the system is not described by a classical field configuration. This can be seen concretely in 
many non-equilibrium quantum field theory techniques, such as the 2PI formalism, 
which describes the dynamics in terms of the correlation functions.
Our results can be employed in such calculations to determine the produced number of defects, or 
to check the validity of these techniques in non-linear situation. 
Indeed, we used this approach in an earlier work~\cite{defects2PI}  to show 
that at next-to-leading order in the 1/N expansion, the 2PI formalism
fails to describe defect formation in 1+1 dimensions.

There is still significant room for improvement in our understanding of the defect signature in
correlation functions. The present results are only valid at relatively low temperature and weak coupling. Thermal and quantum fluctuations change the form of the correlator, both by giving 
a direct contribution to the field correlator and by changing the kink profile~\cite{david},
and more work is needed to disentangle these effects from the contribution due to the kink distribution. Other directions for future work include generalisation to other types of defects, such as vortices and monopoles.

\acknowledgments
A.T. thanks Ossi Tapio for useful discussions. A.R. was supported by STFC. The numerical work was conducted on the Murska Cluster at the Finnish Center for Computational Sciences (CSC).


 
\providecommand{\href}[2]{#2}\begingroup\raggedright

\end{document}